\documentclass[review,number,sort&compress]{elsarticle}

\usepackage{caption}

\usepackage{subfigure} 
\usepackage{rotating}

\usepackage{lineno}  

\usepackage{graphicx}  

\usepackage{xspace}  
\usepackage{color}
\usepackage{colortbl}
\usepackage{rotating}
\usepackage{multirow}
\usepackage{bigints}
\usepackage{amsmath}
\usepackage{ifthen} 
\usepackage[
    plainpages=false,
    bookmarks         = true,
    bookmarksnumbered = true,
    pdfpagemode       = None,
    pdfstartview      = FitH,
    pdfpagelayout     = SinglePage,
    colorlinks        = true,
    linkcolor         = blue,
    pagecolor         = black,
    urlcolor          = blue,
    pdfborder         = {0 0 0}
    ]{hyperref}%
\newboolean{uprightparticles}
\setboolean{uprightparticles}{false} 
\usepackage{amssymb}
\usepackage{amsfonts}
\usepackage{upgreek}
\usepackage{rotating}
\usepackage{multirow}
\usepackage{mciteplus}
\definecolor{miriam}{rgb}{0.65, 0.04, 0.37}
\ifthenelse{\boolean{uprightparticles}}%
{

 \def\PDelta      {\ensuremath{\Delta}\xspace}                 
 \def\PXi      {\ensuremath{\Xi}\xspace}                 
 \def\PLambda      {\ensuremath{\Lambda}\xspace}                 
 \def\PSigma      {\ensuremath{\Sigma}\xspace}                 
 \def\POmega      {\ensuremath{\Omega}\xspace}                 
 \def\PUpsilon      {\ensuremath{\Upsilon}\xspace}                 
 
 \def\PB      {\ensuremath{\mathrm{B}}\xspace}                 
                  
 \def\PD      {\ensuremath{\mathrm{D}}\xspace}

 \def\PK      {\ensuremath{\mathrm{K}}\xspace}

 \def\Pi      {\ensuremath{\mathrm{i}}\xspace}

 \def\Pt      {\ensuremath{\mathrm{t}}\xspace}

}
{

 \mathchardef\PDelta="7101
 \mathchardef\PXi="7104
 \mathchardef\PLambda="7103
 \mathchardef\PSigma="7106
 \mathchardef\POmega="710A
 \mathchardef\PUpsilon="7107
                  
 \def\PB      {\ensuremath{B}\xspace}                 
                  
 \def\PD      {\ensuremath{D}\xspace}

 \def\PK      {\ensuremath{K}\xspace}

 \def\Pi      {\ensuremath{i}\xspace}

 \def\Pt      {\ensuremath{t}\xspace}

}

\def\t     {\ensuremath{\Pt}\xspace}

\def\kaon  {\ensuremath{\PK}\xspace}
  \def\Kbar  {\kern 0.2em\overline{\kern -0.2em \PK}{}\xspace}

\def\Kz    {\ensuremath{\kaon^0}\xspace}
\def\Kzb   {\ensuremath{\Kbar^0}\xspace}
\def\KzKzb {\ensuremath{\Kz \kern -0.16em \Kzb}\xspace}
\def\Kp    {\ensuremath{\kaon^+}\xspace}
\def\Km    {\ensuremath{\kaon^-}\xspace}

\def\KpKm  {\ensuremath{\Kp \kern -0.16em \Km}\xspace}

  \def\Dbar    {\kern 0.2em\overline{\kern -0.2em \PD}{}\xspace}
\def\D       {\ensuremath{\PD}\xspace}

\def\Dz      {\ensuremath{\D^0}\xspace}
\def\Dzb     {\ensuremath{\Dbar^0}\xspace}
\def\DzDzb   {\ensuremath{\Dz {\kern -0.16em \Dzb}}\xspace}
\def\Dp      {\ensuremath{\D^+}\xspace}
\def\Dm      {\ensuremath{\D^-}\xspace}

\def\DpDm    {\ensuremath{\Dp {\kern -0.16em \Dm}}\xspace}

  \def\Bbar    {\kern 0.18em\overline{\kern -0.18em \PB}{}\xspace}

  \def\Y#1S{\ensuremath{\PUpsilon{(#1S)}}\xspace}

\newcommand{\tev}{\ensuremath{\mathrm{\,Te\kern -0.1em V}}\xspace}
\newcommand{\gev}{\ensuremath{\mathrm{\,Ge\kern -0.1em V}}\xspace}
\newcommand{\mev}{\ensuremath{\mathrm{\,Me\kern -0.1em V}}\xspace}
\newcommand{\kev}{\ensuremath{\mathrm{\,ke\kern -0.1em V}}\xspace}
\newcommand{\ev}{\ensuremath{\mathrm{\,e\kern -0.1em V}}\xspace}
\newcommand{\gevc}{\ensuremath{{\mathrm{\,Ge\kern -0.1em V\!/}c}}\xspace}
\newcommand{\mevc}{\ensuremath{{\mathrm{\,Me\kern -0.1em V\!/}c}}\xspace}
\newcommand{\gevcc}{\ensuremath{{\mathrm{\,Ge\kern -0.1em V\!/}c^2}}\xspace}
\newcommand{\gevgevcccc}{\ensuremath{{\mathrm{\,Ge\kern -0.1em V^2\!/}c^4}}\xspace}
\newcommand{\mevcc}{\ensuremath{{\mathrm{\,Me\kern -0.1em V\!/}c^2}}\xspace}

\def\to                 {\ensuremath{\rightarrow}\xspace}

\def\gsim{{~\raise.15em\hbox{$>$}\kern-.85em
          \lower.35em\hbox{$\sim$}~}\xspace}
\def\lsim{{~\raise.15em\hbox{$<$}\kern-.85em
          \lower.35em\hbox{$\sim$}~}\xspace}

\def\AT#1     {\ensuremath{A_T^{#1}}\xspace}           

\def\C#1      {\ensuremath{\mathcal{C}_{#1}}}                       
\def\Cp#1     {\ensuremath{\mathcal{C}_{#1}^{'}}}                    
\def\Ceff#1   {\ensuremath{\mathcal{C}_{#1}^{\mathrm{(eff)}}}}        
\def\Cpeff#1  {\ensuremath{\mathcal{C}_{#1}^{'\mathrm{(eff)}}}}       
\def\Ope#1    {\ensuremath{\mathcal{O}_{#1}}}                       
\def\Opep#1   {\ensuremath{\mathcal{O}_{#1}^{'}}}                    

\newcommand{\CLsb}{\ensuremath{\textrm{CL}_{\textrm{s+b}}}\xspace}
\newcommand{\CLs}{\ensuremath{\textrm{CL}_{\textrm{s}}}\xspace}
\newcommand{\CLb}{\ensuremath{\textrm{CL}_{\textrm{b}}}\xspace}

\newcommand{\pdf}{\ensuremath{p.d.f.}\xspace}

\newcommand{\figref}[1]{Fig.~\ref{#1}}
\newcommand{\tabref}[1]{Table~\ref{#1}}

\newcommand{\secref}[1]{Sect.~\ref{#1}}

\journal{Nuclear Instruments and Methods}

\begin{document}

\begin{frontmatter}
\title{Properties of frequentist confidence levels derivatives}
\author[a]{M. Lucio Mart\'inez}
\author[a]{D. Mart{\'{\i}}nez Santos}
\author[b]{F. Dettori}

\address[a]{Universidade de Santiago de Compostela, E-15706 Santiago de Compostela, Spain}
\address[b]{CERN, CH-1211 Geneva 23, Switzerland}

\begin{abstract}
\noindent In high energy physics, results from searches for new particles or rare processes are often
reported using a modified frequentist approach, known as $\rm{CL_s}$ method. In this paper, 
we study the properties of the derivatives of $\rm{CL_s}$ and $\rm{CL_{s+b}}$ as signal strength estimators if the confidence levels are interpreted 
as credible intervals. Our approach allows obtaining best fit points and $\chi^2$ functions which can be 
used for phenomenology studies. In addition, this approach can be used to incorporate $\rm{CL_s}$ results into
Bayesian combinations.

\end{abstract}



%

\begin{keyword}
statistics \sep limit setting

\end{keyword}

\end{frontmatter}



%



\section{Introduction}
\label{sec:intro}

A method that is commonly used in high energy physics to set limits in production cross sections of hypothetical
new particles~\cite{some_example} as well as in branching fractions of rare decay processes (\emph{e.g.}~\cite{Ksmm,bsmm}), is the so-called \CLs method, or {\it modified frequentist
approach}~\cite{CLs1,CLs2} \footnote{We note that limit setting based on ratio of confidence levels was already used in high energy physics 1973 for rare kaon decays~\cite{Jack}.}. This approach has become very popular because it avoids unphysical limits as well as the possibility
of setting strong limits in experiments with no sensitivity.
However, it does not reveal other potentially relevant information, as for example what is the most probable value for the signal cross section. 
To construct a likelihood function or a probability density function (\pdf, $\rm P(s|data)$)  out of these results, sometimes a Gaussian approach is used. In the Gaussian approach the $90\%~(95\%)$ one-sided limits are converted to 1.6~(2) standard deviations.
However this is in general a very rough approach.

In this paper, we describe a series of methods to obtain posterior probabilities from published $\CLs(s)$ and $\CLsb(s)$ curves. 
These approaches were used for implementing constraints in the phenomenological analyses of Ref.~\cite{nazila,mc75}, as well as to crosscheck the coverage of upper limits based on profile likelihood integration.
The posterior probabilities obtained through these methods can be folded as prior probabilities for Bayesian combinations
with other results.

\clearpage
\newpage

\section{Mathematical definitions}
\label{sec:formulas}

Our goal is to obtain a \pdf  for the signal strength such that the credible intervals obtained with this function match the frequentist confidence levels \CLsb or the confidence level ratio \CLs given by an experiment.
{Hereafter we restrict ourselves to a single signal-strength parameter $s$. The quantities \CLsb and \CLs are assumed to be monotonically decreasing with 
$s$, as expected from any routine meant to set upper limits. Continuity and differentiability are desired, although not strict requirements since
in practice the derivatives will be approximated numerically.
Because the credible interval (or Bayesian confidence level) is the integral of the \pdf, the quantity}:
\begin{equation}
\label{eq:def_delta}
\delta(s) \equiv - \frac{d\CLsb}{ds} \quad , 
\end{equation}
\noindent where $s$ represents the signal strength, gives us a \pdf with credible intervals that are equivalent to the frequentist \CLsb and that therefore has by construction
the appropriate coverage:
\begin{equation}
\label{eq:int_delta}
\CLsb(\sigma) = \int_{-\sigma_{min}}^{\sigma}\delta(s)ds
\end{equation}
\noindent where $\sigma_{min}$ is the minimum value of the signal strenght (including negative values) which is still consistent with non-negative
entries in all the bins, so that Poisson statistics still applies. Arbitrary integration constants have been omitted. 

However, upper limits are commonly set using \CLs and not \CLsb. On one hand, this sacrifices part of the coverage, but on the other hand avoids excluding the null hypothesis as well as obtaining strong limits in experiments with no sensitivity. 
Therefore we define the quantity:
\begin{equation}
\label{eq:def_phi}
\phi(s) \equiv - \frac{d\CLs}{ds}
\end{equation}
\noindent to provide the \pdf with credible intervals equivalent to \CLs limits: 
\begin{equation}
\label{eq:int_phi}
\CLs(\sigma) = \int_{0}^{\sigma}\phi(s)ds\quad . 
\end{equation}
In the particular case of a single bin analysis and without systematic uncertainties,
$\phi(s)$ is equivalent to a posterior built from the likelihood function multiplied by a constant positive $s$ prior~\cite{CLs2}.
Normalization constants should be set such that $\delta(s)$ and $\phi(s)$ integrate to unity over the s domain. 

The function $\delta(s)$ is closely related to the likelihood function. Indeed, as discussed in Appendix 7 of~\cite{Joel}, Bayesian
credible intervals have the frequentist coverage averaged with respect to the prior density. As $\delta(s)$ has by construction the frequentist coverage, it is expected that the corresponding prior that weighs the coverage is constant or nearly constant in the entire phase space.
However, note that $\delta(s)$ can differ from the profile likelihood, due to the fact that credible integrals of the latter do not always have the required coverage, while credible intervals of $\delta(s)$ do.

In the case of a single bin experiment it is easy to demonstrate that $\delta(s)$ corresponds to the likelihood function. Since the following relation holds
\begin{equation}
-\frac{d\CLsb}{ds} = -\sum_{N=0}^{N_{obs}}\frac{ df(s, N)}{ds}\quad , 
\end{equation}
where $N$ is the number of events\footnote{For a single bin experiment $N$ is an optimal test-statistic.} and $f(s ,N)$ is the two-dimensional distribution of the possible outcomes of the test-statistic in the $s,N$ plane, given by Poisson statistics as
\begin{equation}
f(s,N) = \frac{(s+b)^{N}e^{-(s+b)}}{N!} \quad,
\end{equation}
then
\begin{eqnarray}
\delta(s)_{count} &=& -\frac{d\CLsb}{ds}_{count}=  -\sum_{N=0}^{N_{obs}}\frac{ df(s, N)}{ds} = \nonumber \\
&=& \sum_{N=0}^{N_{obs}}\left( \frac{(s+b)^{N}e^{-(s+b)}}{N!} -\frac{N(s+b)^{N-1}e^{-(s+b)}}{N!} \right) \\
&=& \frac{(s+b)^{N_{obs}}e^{-(s+b)}}{N_{obs}!} = f(N_{obs} | s)_{count} \nonumber\quad .
\end{eqnarray}
One property of $\delta(s)$, independent of the choice of test-statistic, is such that if multiplied by a constant positive $s$ prior $\theta(s)$ one finds that:
\begin{multline}
\label{eq:ts}
\frac{\bigintsss_{-\infty}^{\sigma} \theta(s)\delta(s)ds}{\bigintsss_{-\infty}^{+\infty} \theta(s)\delta(s)ds} = \frac{\bigintsss_{\lim_{s\to0}s}^{\sigma}\delta(s)ds} {  \bigintsss_{\lim_{s\to0}s}^{+\infty}\delta(s)ds  } = 
\frac{\lim_{s\to0}\CLsb(s)-\CLsb(\sigma)}{\lim_{s\to0}\CLsb(s)-0}  = \\
= \frac{\lim_{s\to0}\CLb(s)-\CLsb(\sigma)}{\lim_{s\to0}\CLb(s)} \approx 1 - \CLs(\sigma)
\end{multline}
\noindent i.e. upper limits derived from the integration of $\delta(s)$ on the positive range of $s$ are expected to be very close to those obtained
from \CLs. The above approximation is exact in the case where \CLb is independent of $s$, which is possible for certain choices of the test-statistic. In these cases, $\theta(s)\delta(s) = \phi(s)$.
This interesting relation leads us to define:
\begin{equation}
\epsilon(s) = \frac{\phi(s)}{\delta(s)}
\end{equation}
\noindent which can be understood as the effective prior on $s$ needed to get limits equivalent to \CLs.

It can also be noticed that:
\begin{equation}
\label{eq:p_value}
1- \int_0^{+\infty}\delta(s) ds = 1 - \lim_{s\to 0} \CLb(s)
\end{equation}
which can be used to build background-only $p$-values. This is interesting as from a combination of null searches, with only upper limits, a non-null combination could also be derived. In particular, for a test-statistic in which $\CLb$ does not depend on $s$,
the usual $p$-value used in \CLs should be identical to what is obtained from Eq. \eqref{eq:p_value}. 
On the contrary, $\phi(s)$ is by construction normalized to 1 in the positive range of $s$.
When $\phi$ or $\delta$ are interpreted as probabilities, one can build a $\chi^2$ function out of them as:
\begin{eqnarray}
\label{eq:chi2}
\chi^2_{\delta} (s) &=& -2 \log{\frac{\delta(s)}{ max\left\{\delta(s)\right\}}}\\
\chi^2_{\phi} (s) &=& -2 \log{\frac{\phi(s)}{ max\left\{\phi(s)\right\}}}\quad .
\end{eqnarray}
 
\clearpage
\newpage

\section{Numerical tests and examples}
\label{sec:tests}

In this section we demonstrate few examples of $\delta(s)$, $\phi(s)$ and $\epsilon(s)$ functions as obtained from certain hypothetical experiments.

\subsection{Classic case}
\label{ssec:classic}

We start by showing examples on how $\delta(s)$, $\phi(s)$ and $\varepsilon(s)$ behave when no
systematic uncertainties are included. A typical test-statistic used in the \CLs method is: 
\begin{equation}
Q = \prod_i \frac{ e^{-(s_i+b_i)}(s_i+b_i)^{d_i} / d_i!} {  e^{-(b_i)}(b_i)^{d_i} / d_i! }
\end{equation}
where {$s_i$ and $b_i$ represent the number of expected signal and background events in the \textit{i}-th bin, respectively, and $d_i$ refers to the number of observed events in the same bin.} 
We use the \texttt{mc\_limit} package~\cite{mc_limit} for calculating \CLs and \CLsb in the examples of this subsection.

\begin{figure} [ht]
\begin{center}
\includegraphics[scale=0.30]{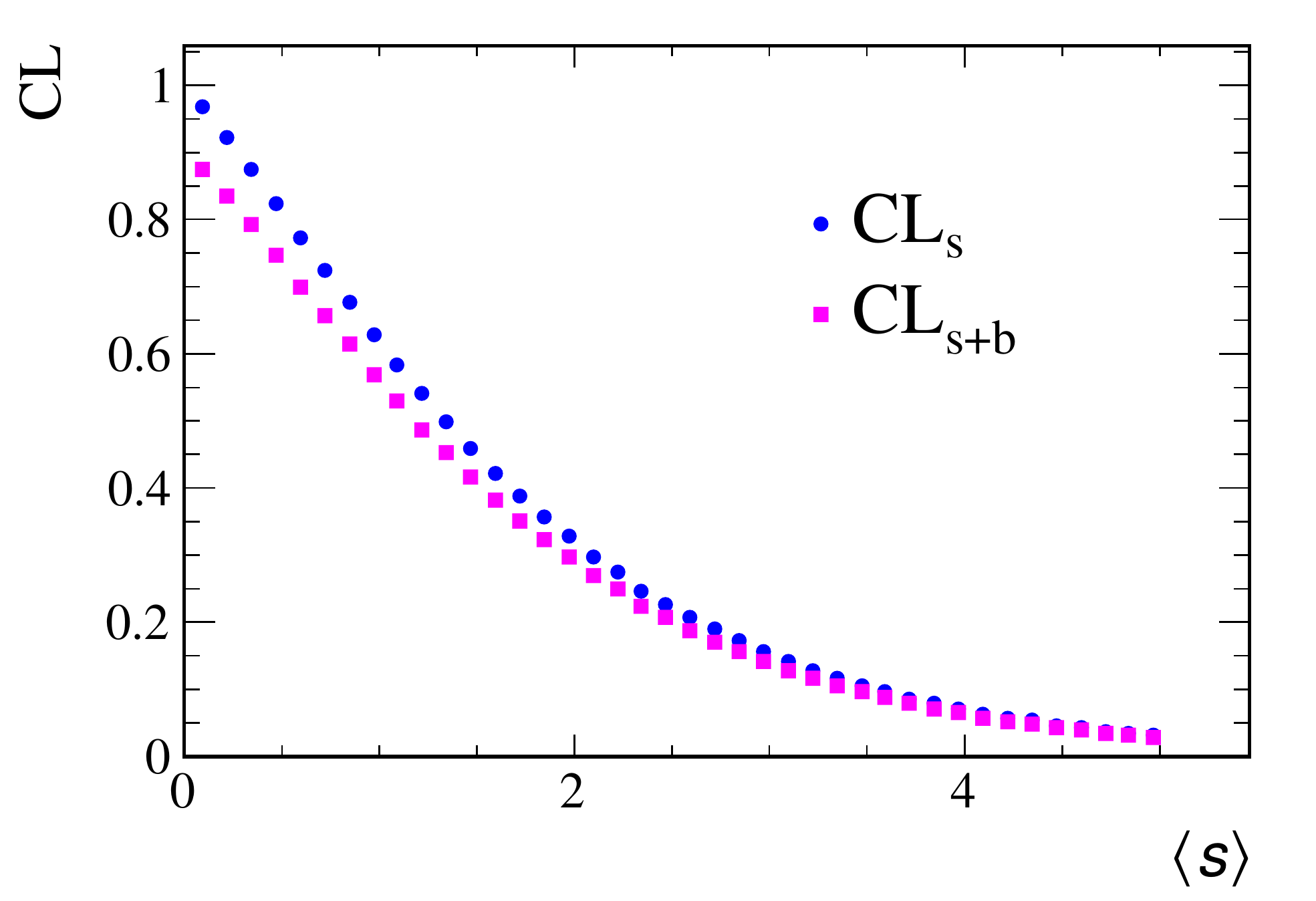} 
\caption{\CLs (blue circles) and \CLsb (magenta squares) as a function of the signal expectation, $\left \langle s \right \rangle$, calculated using 600k toy experiments.\label{fig:CL}}
\end{center}
\end{figure}

\begin{figure} [htb!]
\begin{center}
\includegraphics[scale=0.30]{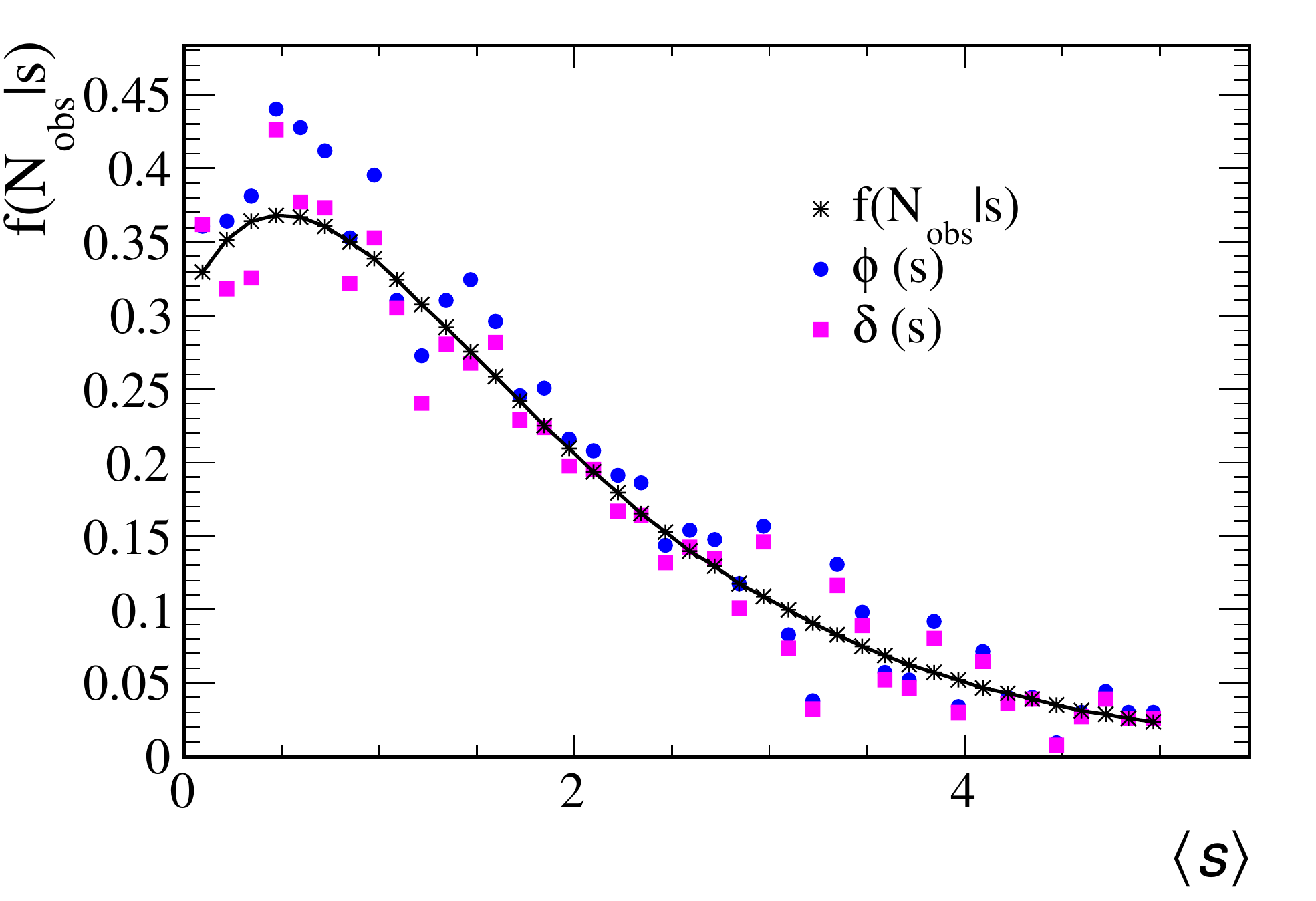} 
\includegraphics[scale=0.30]{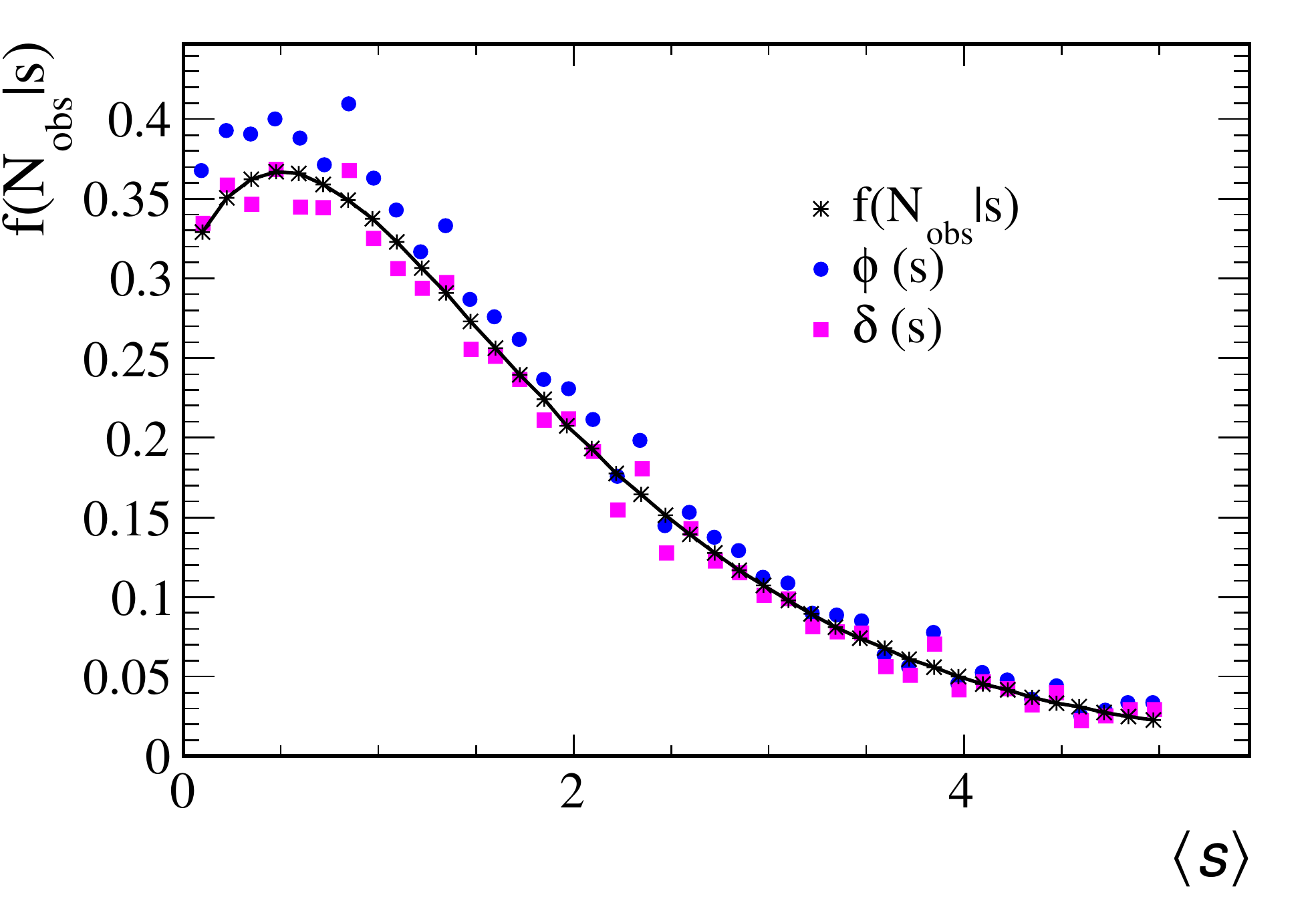} 
\includegraphics[scale=0.30]{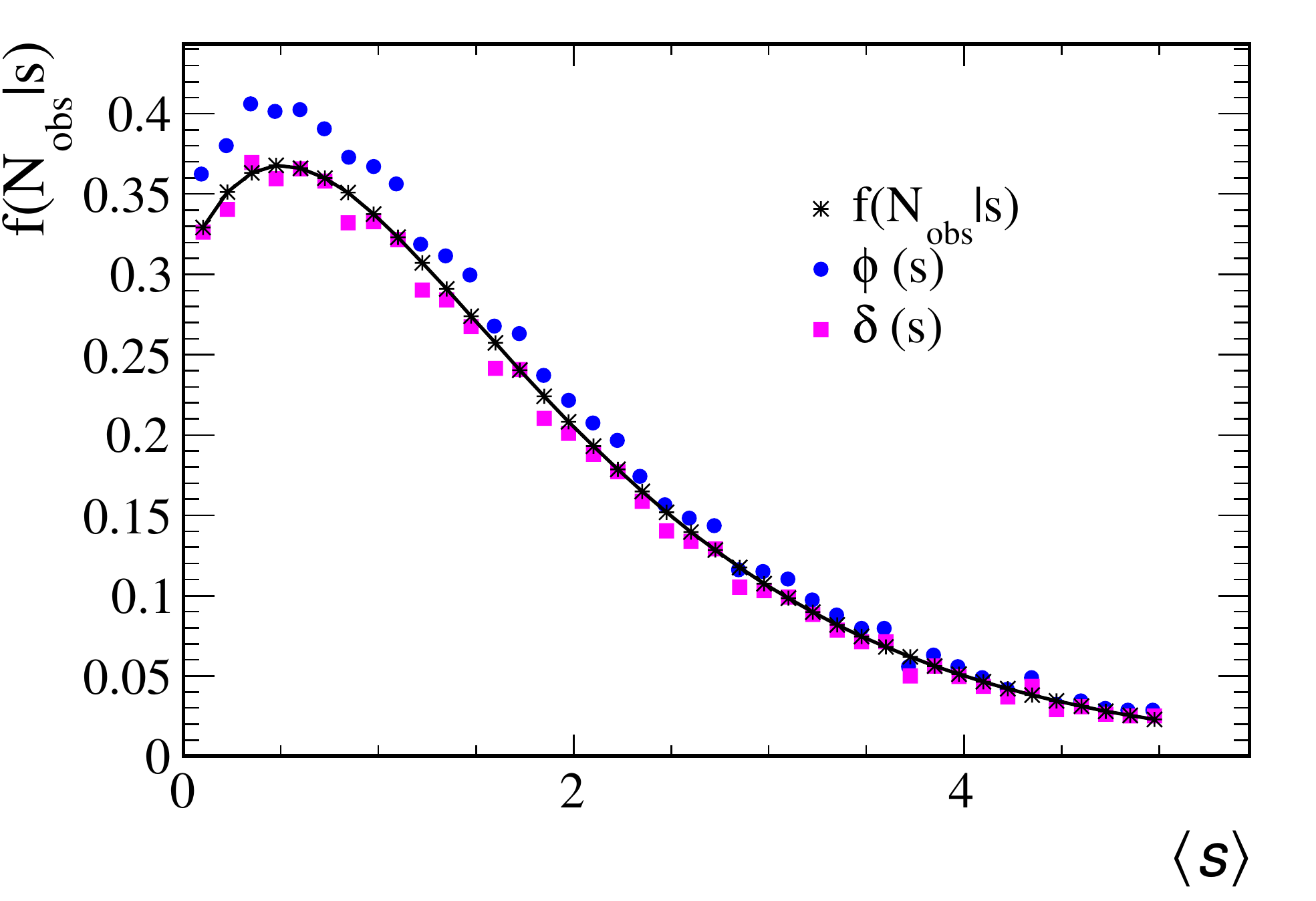} 
\includegraphics[scale=0.30]{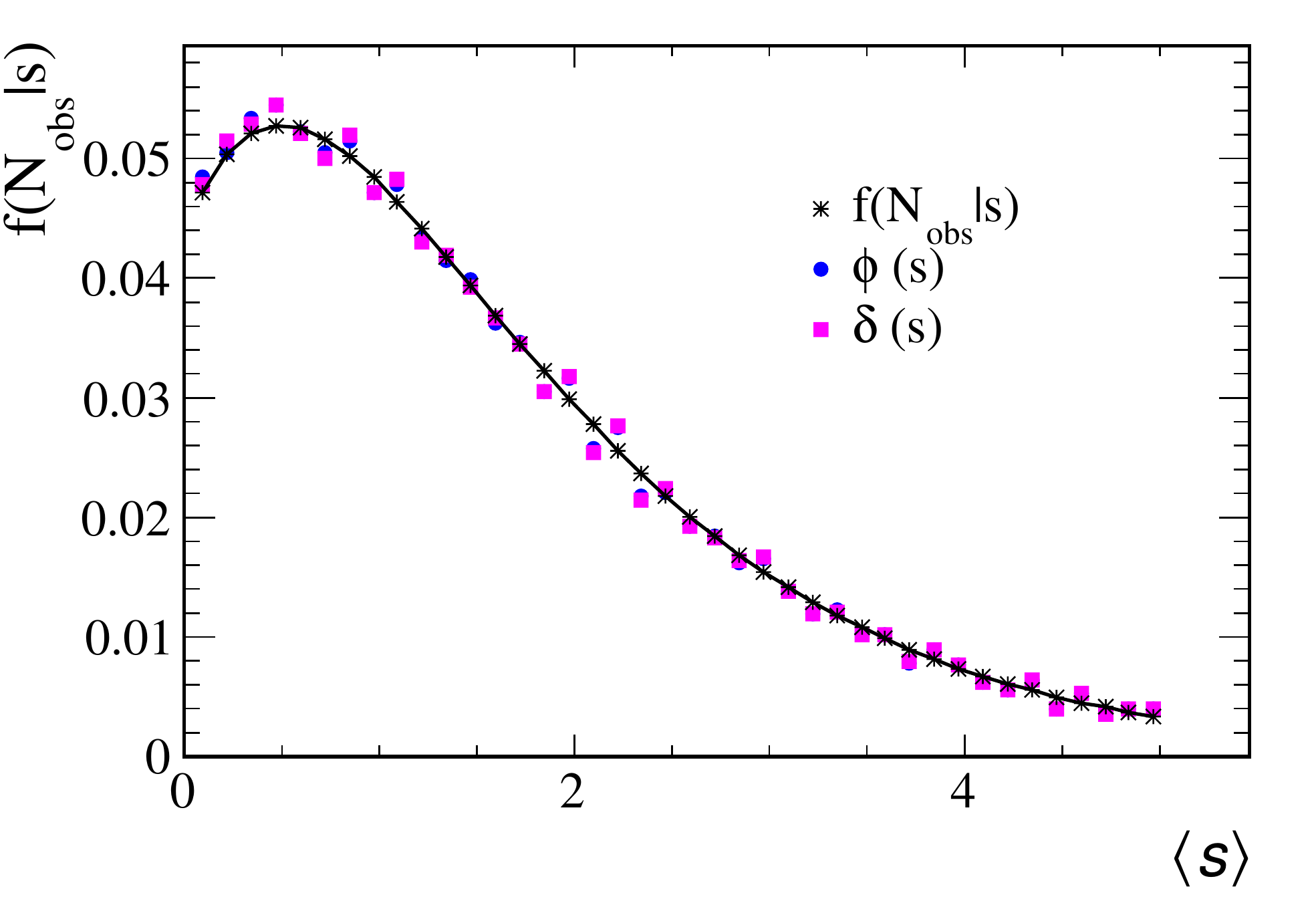} 
\caption{$f(N_{obs} | s)_{count}$ (black solid line), $\delta(s)$ (blue circles), and $\phi(s)$ (magenta squares) calculated using 20k (upper left), 200k (upper right) and 600k (bottom left) toy experiments. The plot on the bottom right shows the same curves as in bottom left, but all normalized to have the same area (i.e, normalized in the $s>0$ range). \label{fig:dCL}}
\end{center}
\end{figure}

\begin{figure} [htb!]
\begin{center}
\includegraphics[scale=0.30]{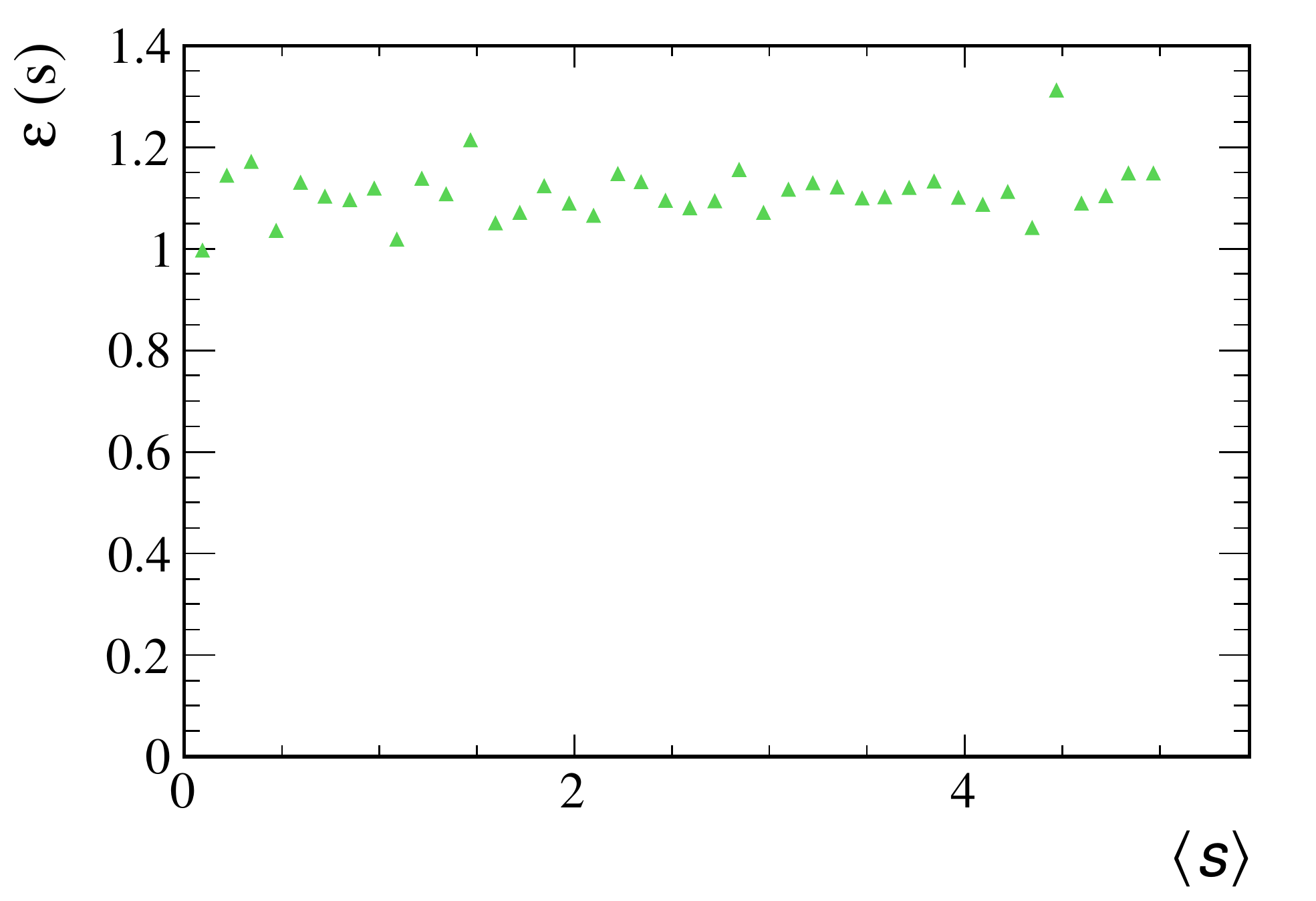} 
\includegraphics[scale=0.30]{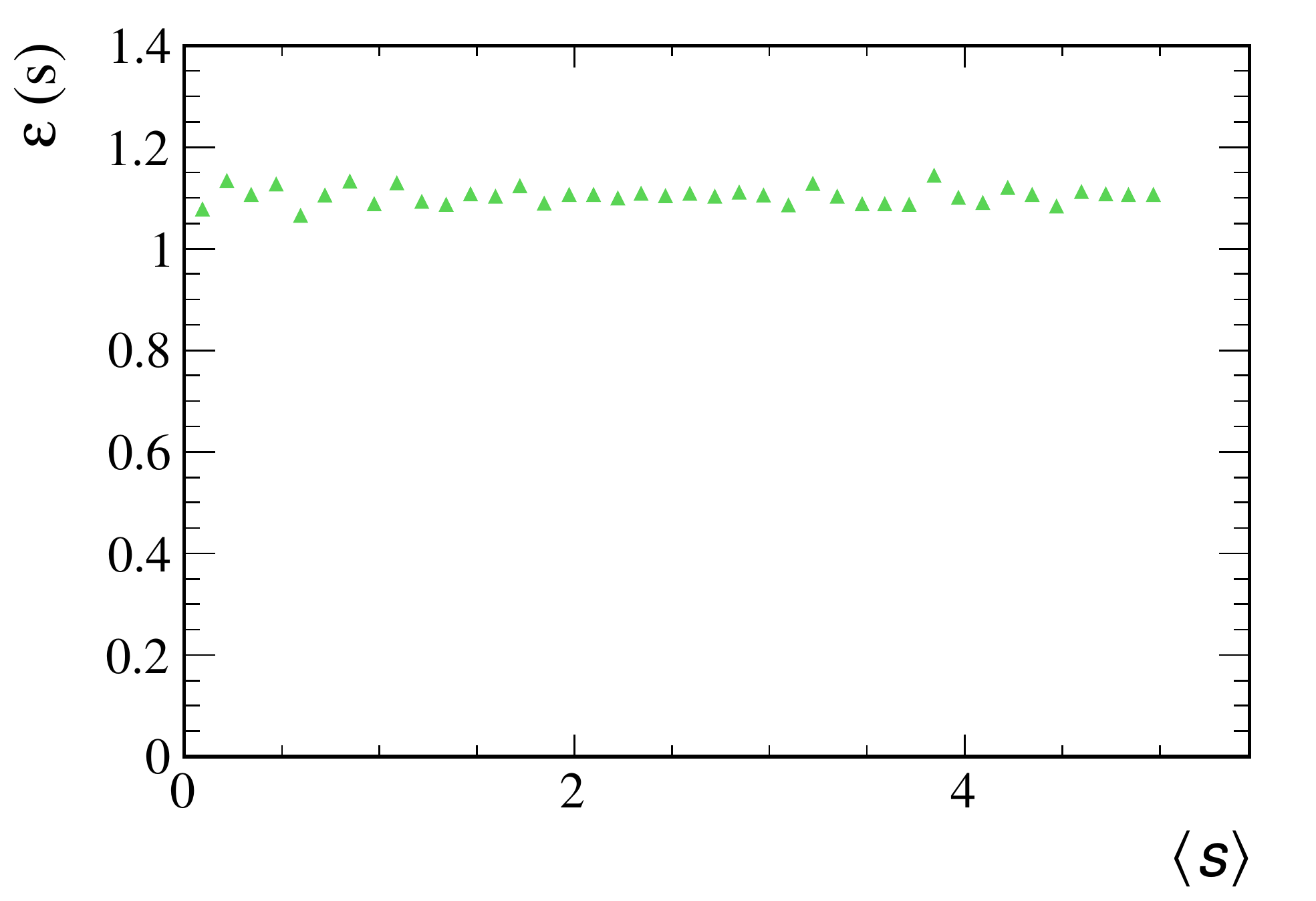} 
\includegraphics[scale=0.30]{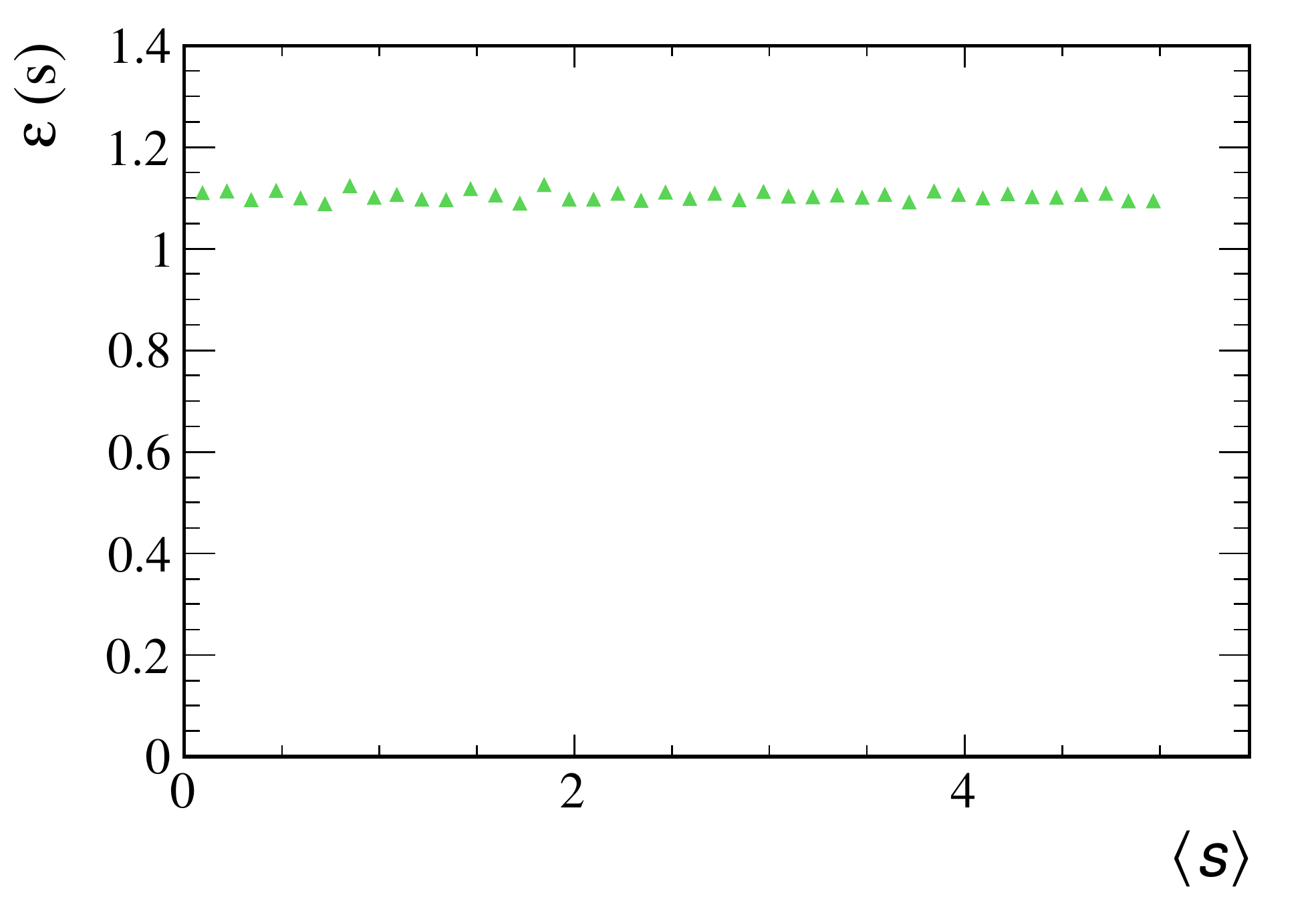} 
\caption{$\varepsilon(s)$ for 20k (upper left), 200k (upper right) and 600k (center) toy experiments.
\label{fig:eps} }
\end{center}
\end{figure}

In our tests, we calculate the derivatives in a numeric way to obtain $\delta$ and $\phi$.
In order to get a smooth lineshape for $\delta(s)$ one needs to run a large amount of toy experiments,
to generate values of \CLsb with enough digits. \figref{fig:CL} shows both \CLs and \CLsb as a function of the signal strength for a single bin experiment, with a background expectation of 0.5 events, a signal expectation varying between 0 and 5 events, and an observation of one event.
For each value of $ s $, $f(N_{obs} | s)_{count}$ is shown, and compared to the $\delta(s)$ and $\phi(s)$ functions calculated using 20k, 200k and 600k pseudo-experiments (\figref{fig:dCL}). The ratio between $\phi$ and $\delta$,   $\varepsilon(s)$, is computed and shown for the different sets of pseudo-experiments (\figref{fig:eps}). It can be seen that a large number of pseudo-experiments is needed in order to properly recover the shape of $\delta$, $\phi$, and $\varepsilon$.

\subsection{Difference between two fits as test-statistic}
\label{ssec:DLL}
\begin{figure} [htb!]
\begin{center}
\includegraphics[scale=0.50]{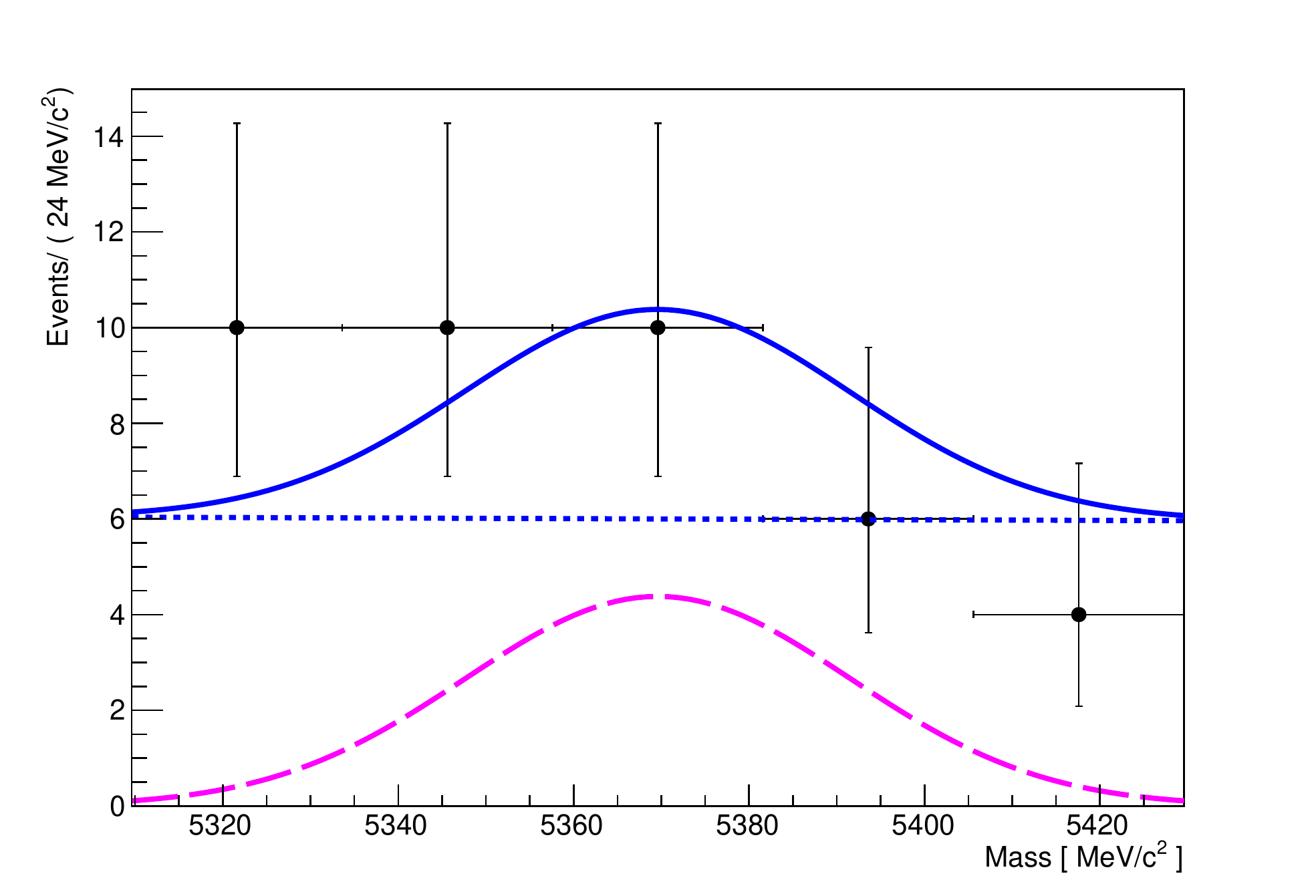}
\caption{{Mass distribution of the generated data for the experiments with (\secref{ssec:sys}) and without (\secref{ssec:DLL}) systematic uncertainties. The magenta dashed line shows the signal contribution, the dotted blue line the background, and the solid blue line the total model.}\label{fig:DLL_exp1}}
\end{center}
\end{figure}

In this example we will explore the use of a test-statistic constructed as the likelihood ratio between
the background hypothesis and the best fit point for $s$, hereafter $\hat{s}$:
\begin{equation}
R = \prod_i \frac{ e^{-(\hat{s}f_i+b_i)}(\hat{s}f_i+b_i)^{d_i} / d_i!} {  e^{-(b_i)}(b_i)^{d_i} / d_i! }
\end{equation}
\noindent where $f_i$ is the signal fraction in the $i$-th bin, so that $\Sigma f_i = 1$. The test statistic
$R$ is independent of the signal hypothesis and so it will be \CLb. Therefore, in this case Eq. \eqref{eq:ts} should hold exactly.
As a numerical example we will use a search experiment of a gaussian signal in the
mass spectrum, on top of an exponential background. The mass range is [5309.6, 5429.6] \mevcc, and the mass
peak is at 5369.6 \mevcc with a resolution of 22 \mevcc. The background slope is assumed to be $-10^{-4}/(\mevcc)$.
In \figref{fig:DLL_exp1} we show the mass distribution of the generated data, superimposed with the best fit. In \tabref{tab:cl} we show the 90 and 95\% upper limits from \CLs and \CLsb obtained from those curves, {using both R and Q as test-statistics}. 

\begin{table}[htb!]
\begin{center}
\caption{90\% and 95\% upper limits on $s$ as obtained with different estimators and test statistics (See text for details).}
\label{tab:cl}
\begin{tabular}{r|c|c}
 Confidence Level & Estimator and test statistic & Excluded $s$ \\
 \hline
90\% &$\CLs^Q$,$\CLs^R$ & 16.9, 17.0\\ 
90\% &$\CLsb^Q$,$\CLsb^R$ & 16.8, 16.8 \\ 
95\% &$\CLs^Q$,$\CLs^R$ &  19.3, 19.3\\ 
95\% &$\CLsb^Q$,$\CLsb^R$ & 19.2, 19.1 \\ 
 \hline
\end{tabular}
\end{center}
\end{table}

\begin{figure} [htb!]
\begin{center}	
\includegraphics[scale=0.40]{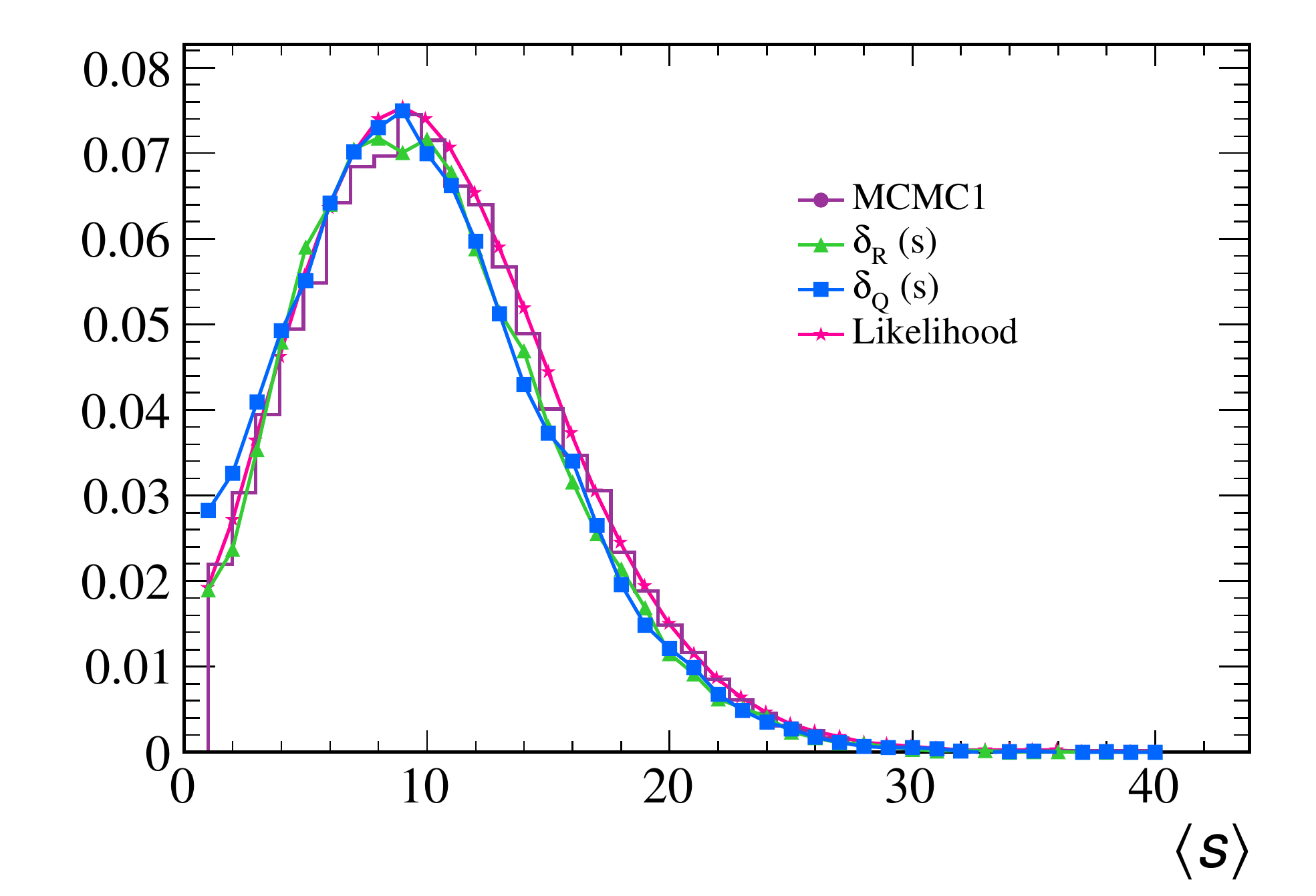}
\includegraphics[scale=0.40]{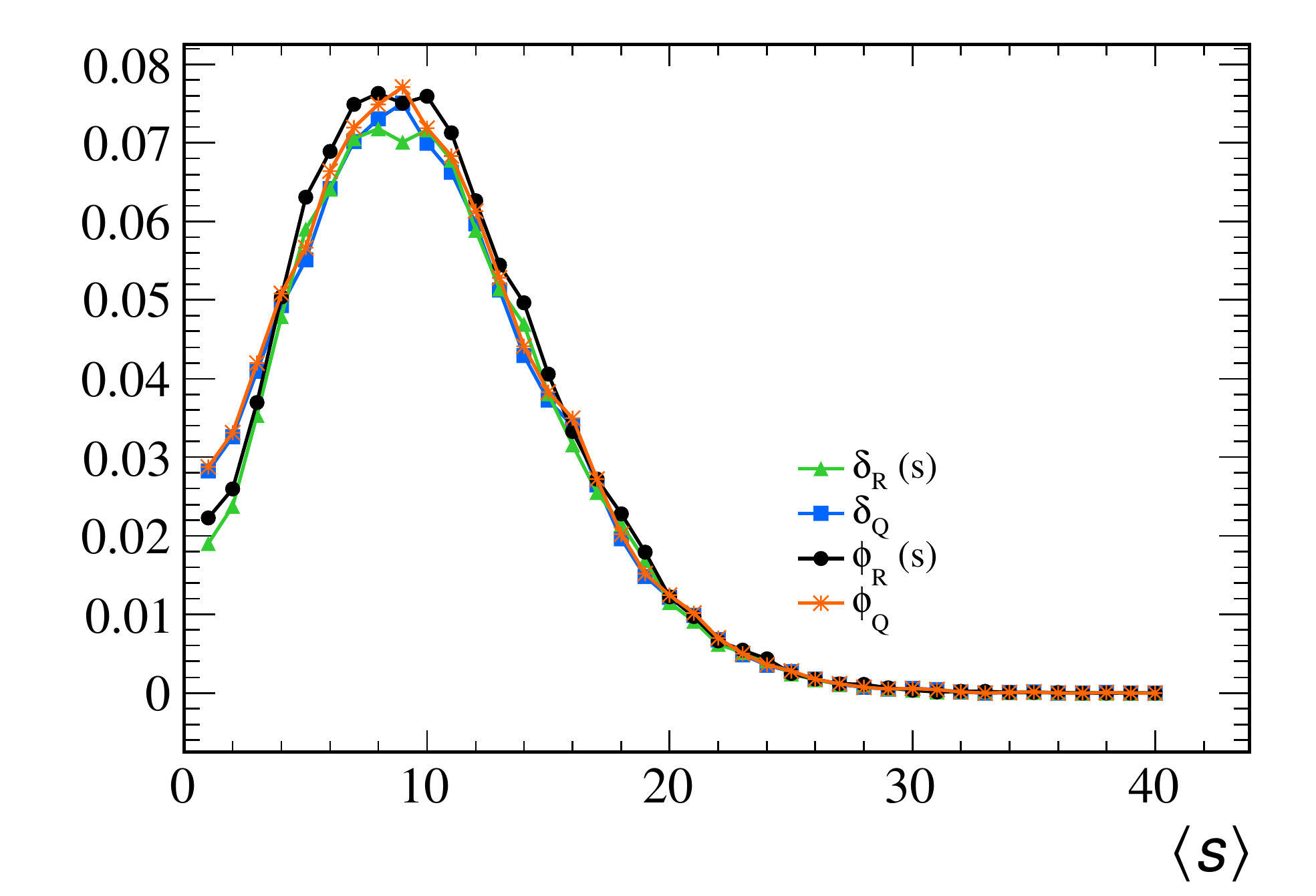}
\includegraphics[scale=0.40]{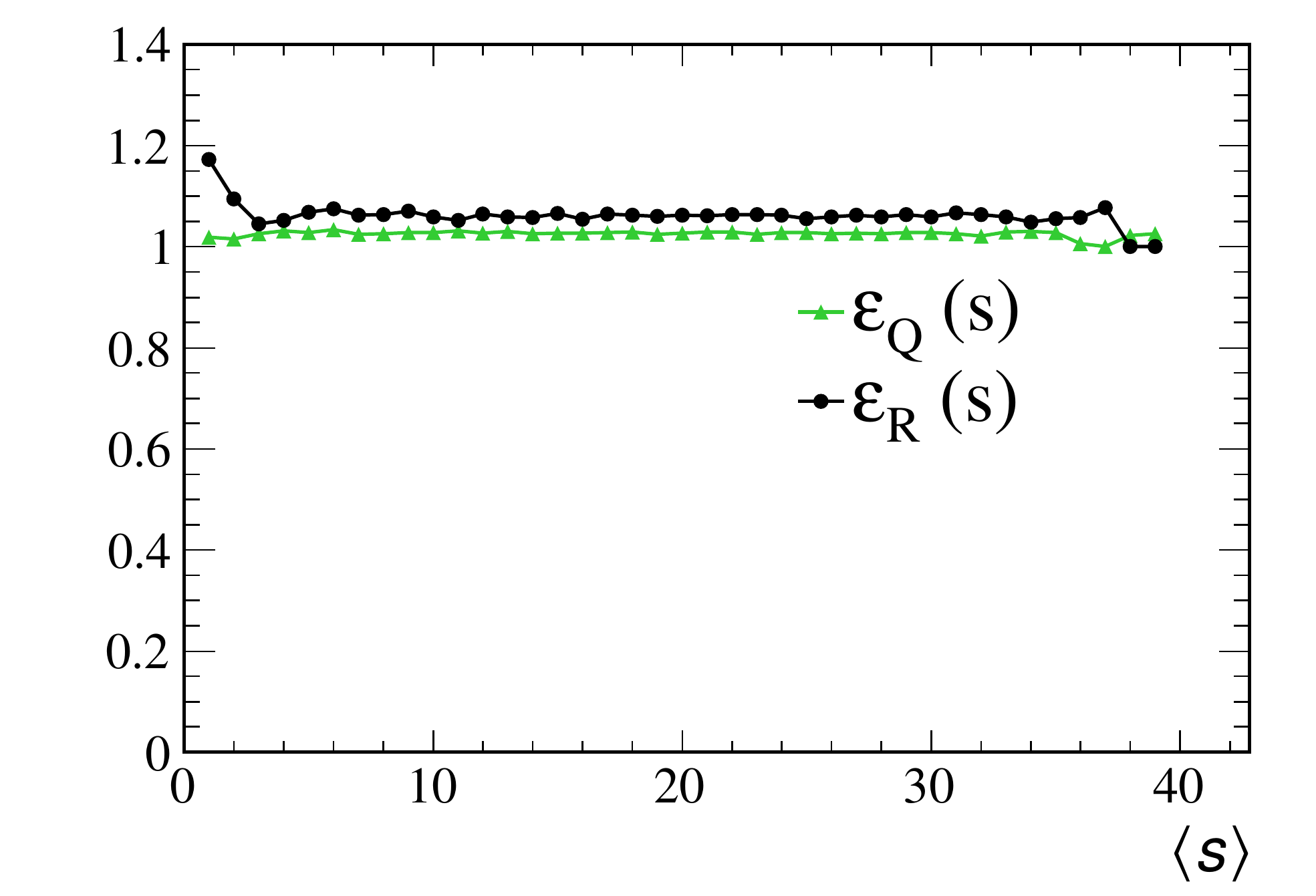}
\caption{{Top: \texttt{mcmc1} posterior from \texttt{mc\_limit} (violet solid line), $\delta_R(s)$ (green triangles), $\delta_Q(s)$ (cyan squares), likelihood (pink stars). Medium: $\delta_R(s)$ (green triangles), $\delta_Q(s)$ (cyan squares), $\phi_R(s)$ (black circles), $\phi_Q(s)$ (orange asterisks). Bottom: $\varepsilon_Q(s)$ (green triangles), $\varepsilon_R(s)$ (black circles). Experiment with no systematics (\secref{ssec:DLL}).}\label{fig:DLL_deltas}}
\end{center}
\end{figure}

In \figref{fig:DLL_deltas} we show the $\delta(s)$, $\phi(s)$ and $\epsilon(s)$ curves obtained for the two choices of
test-statistic, and compared with the likelihood scan from a fit to the data, as well as with the Markov Chain Monte Carlo posterior \texttt{mcmc1} obtained from \texttt{mc\_limit}~\cite{mc_limit}.  It can be seen that $\delta(s)$
is very similar to the likelihood function, and $\phi(s)$ to the likelihood multiplied by a flat $s>0$ prior. 

\subsection{Fit with systematic uncertainties}
\label{ssec:sys}

In the following example we will use a search experiment of a gaussian signal in the
mass spectrum, on top of an exponential background, as it is done in \secref{ssec:DLL}, but
in the presence of the following nuisance parameters:

\begin{itemize}
\item $N_b$, the expected number of background events, which is approximately known
from an external source.
\item $\kappa$, the coefficient of the exponential mass \pdf of the background.
\item $M$, the peak position.
\item $\sigma$, the invariant mass resolution.
\end{itemize}

\begin{figure} [htb!]
\begin{center}
\includegraphics[scale=0.40]{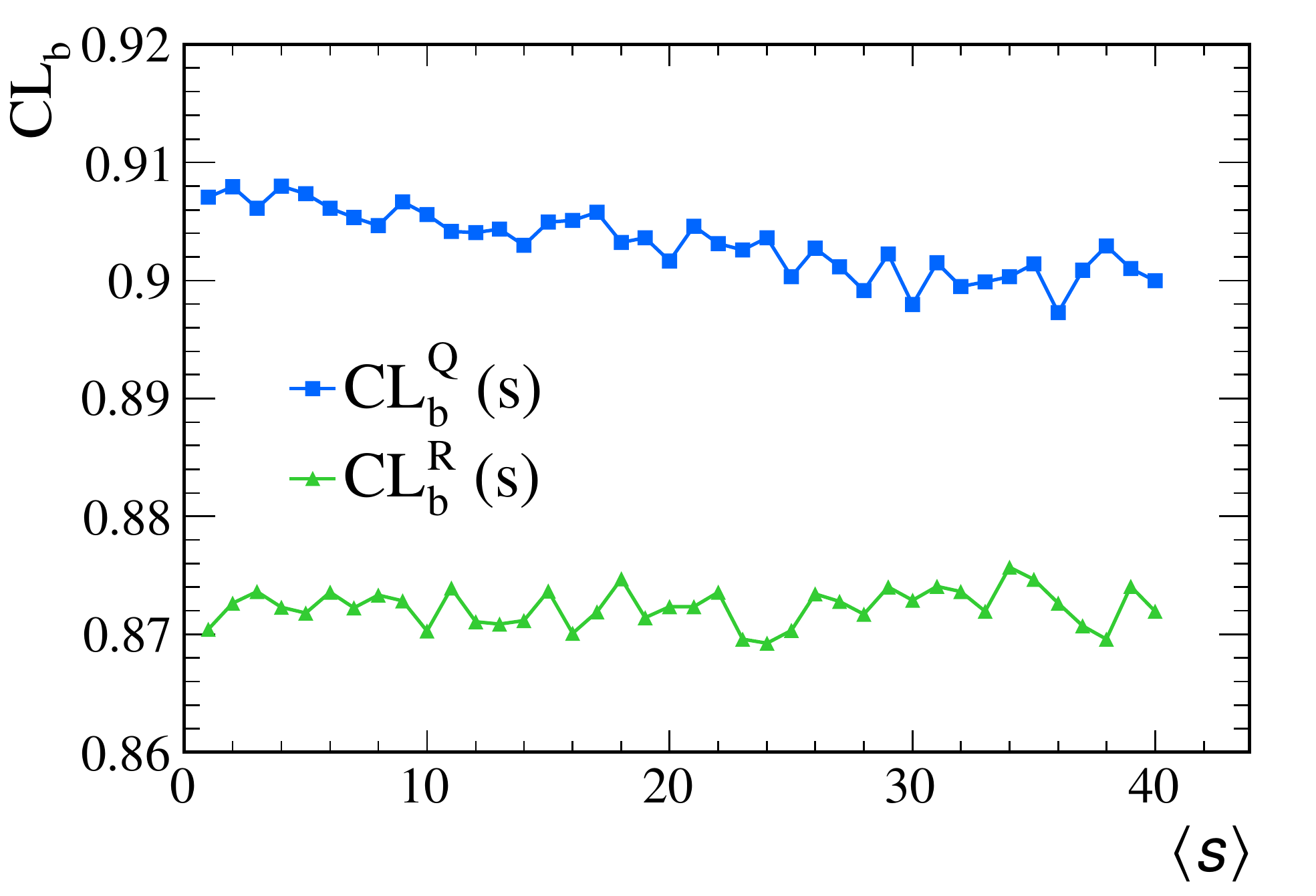}
\caption{{$\CLb^Q$ (cyan squares) and $\CLb^R$ (green triangles) as a function of the signal expectation, $\left \langle s \right \rangle$. Experiment with no systematics (\secref{ssec:DLL}).}
\label{fig:clb}}
\end{center}
\end{figure}

\begin{figure} [H]
\begin{center}
\includegraphics[scale=0.40]{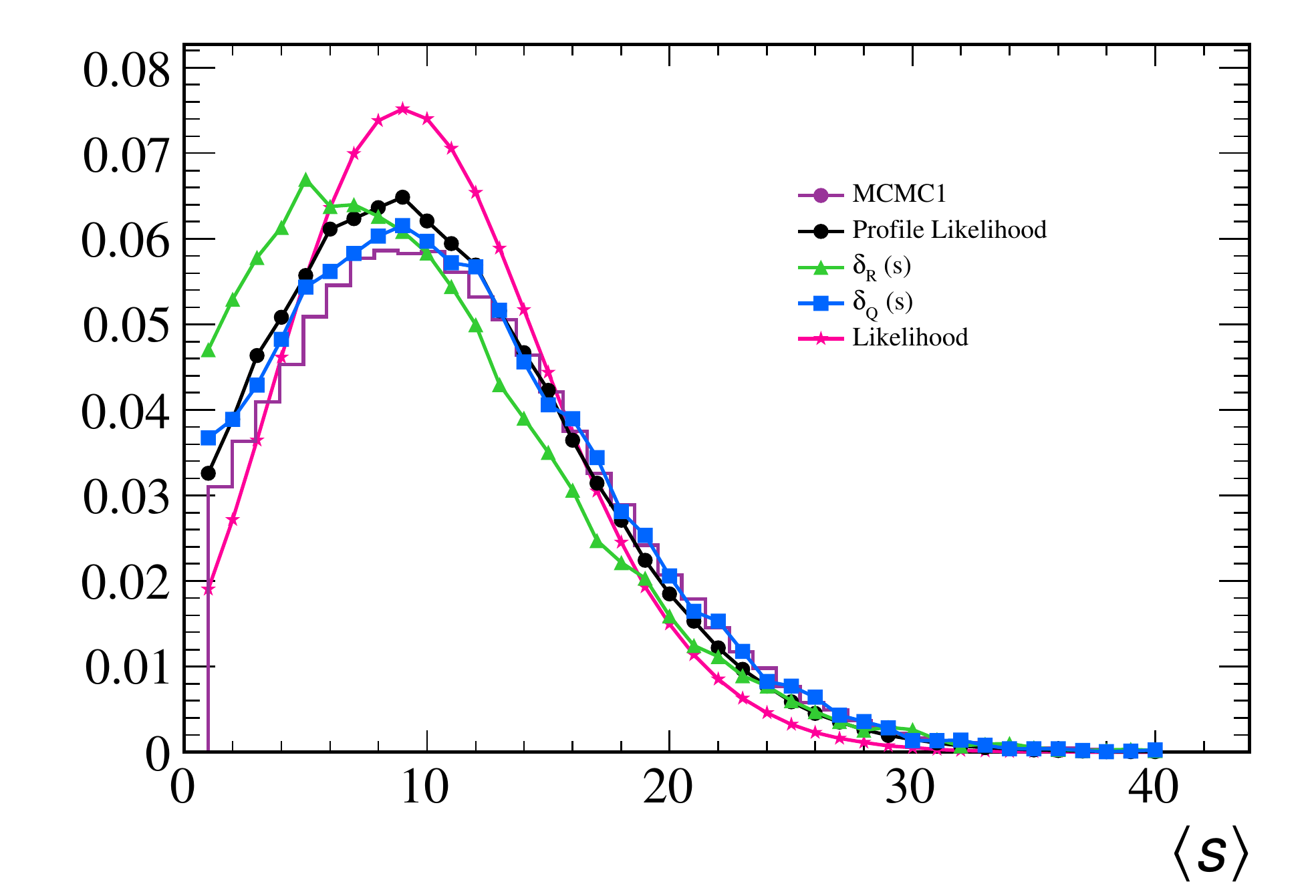}
\includegraphics[scale=0.40]{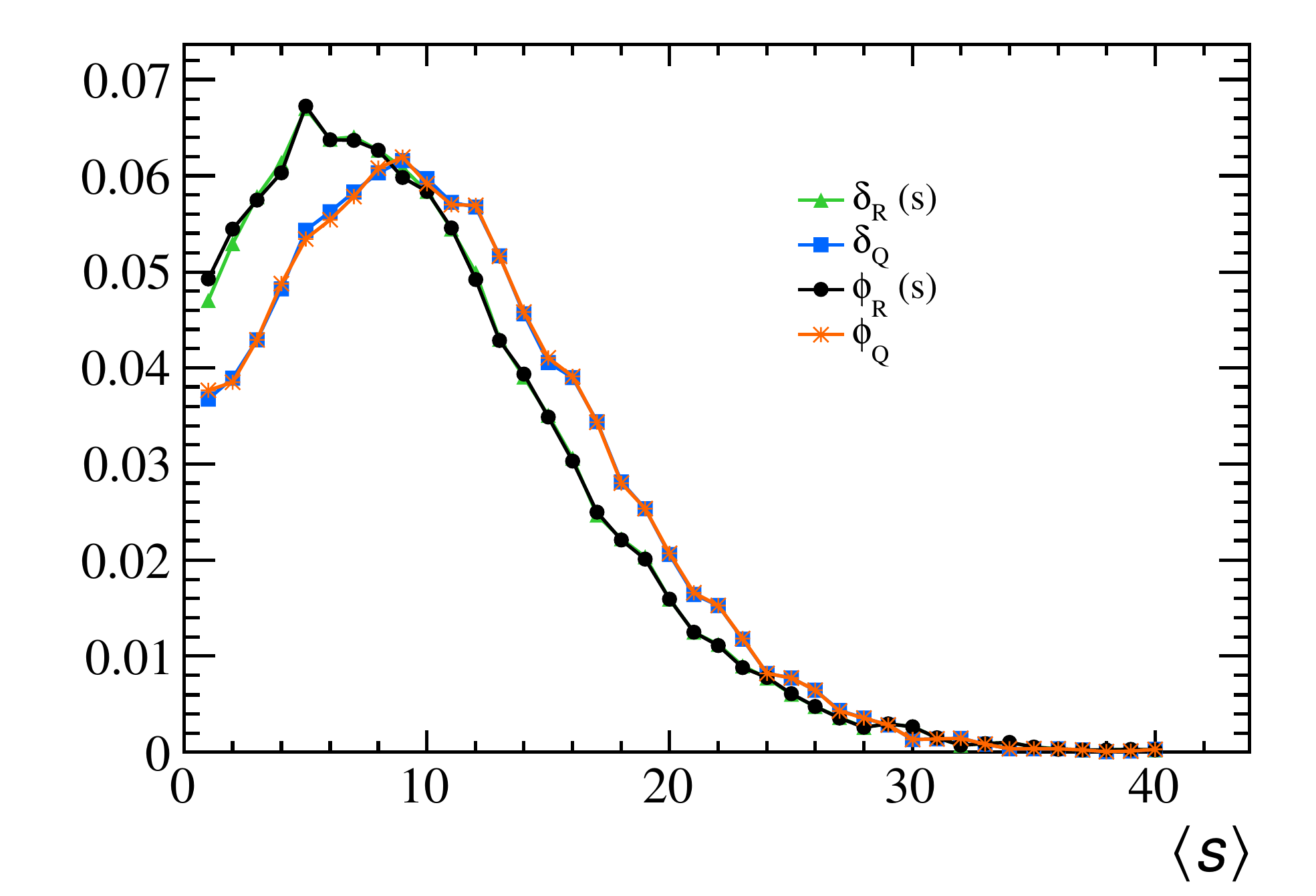}
\includegraphics[scale=0.40]{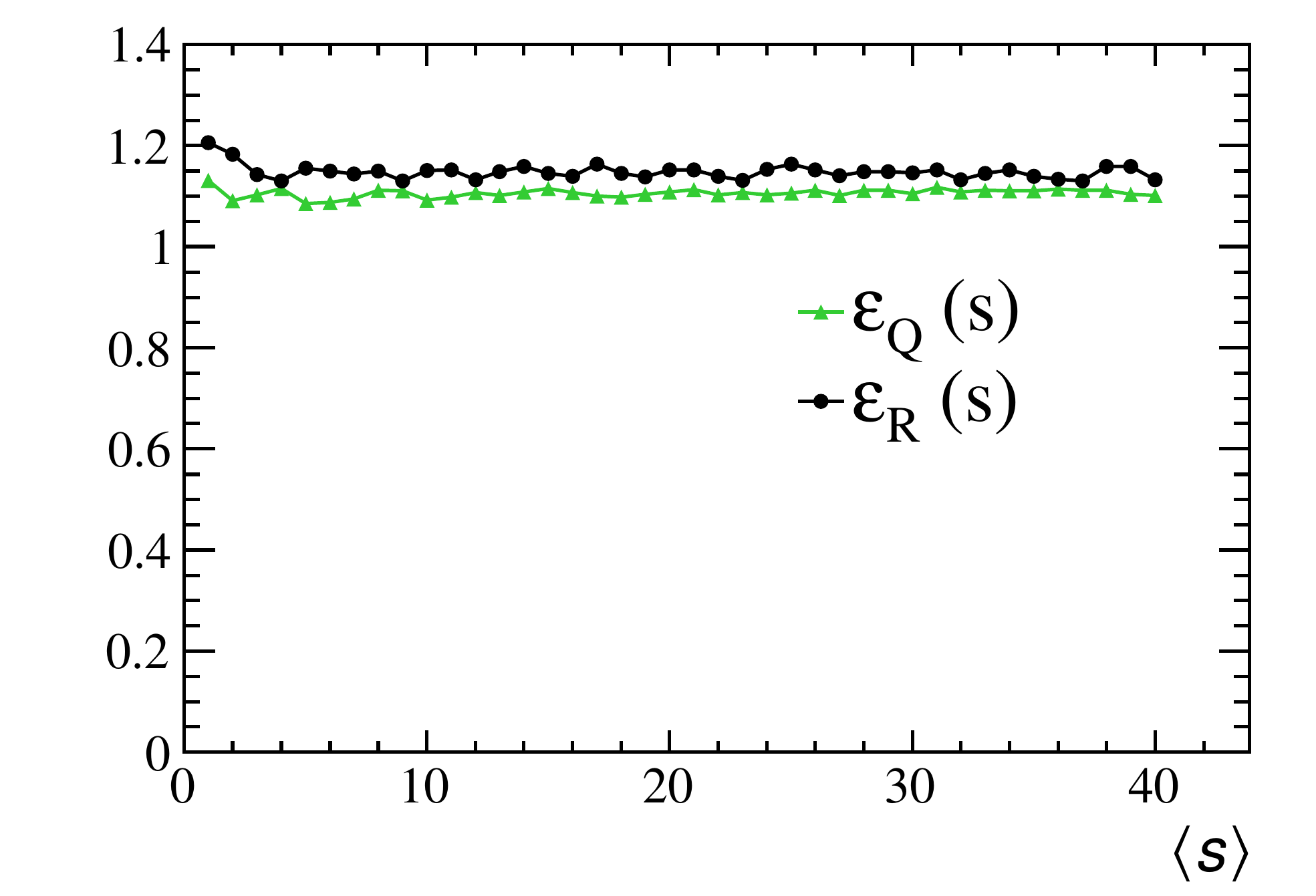}
\caption{{Top: MCM1 (violet solid line), profile likelihood (black circles), $\delta_R(s)$ (green triangles), $\delta_Q(s)$ (cyan squares), likelihood (pink stars). Medium: $\delta_R(s)$ (green triangles), $\delta_Q(s)$ (cyan squares), $\phi_R(s)$ (black circles), $\phi_Q(s)$ (orange asterisks). Bottom: $\varepsilon_Q(s)$ (green triangles), $\varepsilon_R(s)$ (black circles). Experiment with systematics (\secref{ssec:sys}), where $N_{obs} > \left \langle b \right \rangle$.}
\label{fig:dclsys}}
\end{center}
\end{figure}

{All these parameters are considered to have gaussian errors.}
The test statistic will be the difference in the log-likelihood between a fit with the signal strength set to zero and a fit
with the signal strength free. The nuisance parameters are fitted taking into account their prior constraints.
The ensembles of pseudo-experiments are generated fluctuating the nuisance parameters according to their prior probabilities.
In \figref{fig:clb} we compare the \CLsb and \CLb curves obtained using $Q$ and $R$ test-statistics. We see that, as expected, $\CLb^R$ is independent of
$s$, which is not the case of $\CLb^Q$. This is a very useful property, since otherwise one has to make a choice of 
$s$ in a somewhat arbitrary manner in order to report a background $p$-value.
In \figref{fig:dclsys} we show {$\delta(s)$, $\phi(s)$ and $\epsilon(s)$} using the nuisance parameter values as listed in \tabref{tab:np_1}. 

\begin{table}[htb!]
\caption{Nuisance parameter values for the numeric example}
\label{tab:np_1}
\begin{center}
\begin{tabular}{c|c}
 Nuisance Parameter & value \\
 \hline
$\kappa$ & $(-1.0\pm0.3)\times 10^{-4} (\mevcc)^{-1}$ \\
$M$ & $ 5369.6 \pm 1 \mevcc$ \\
$\sigma$ & $22 \pm 2 \mevcc$ \\
$N_b$ & $30_{-5}^{+15}$ \\
 \hline
\end{tabular}
\end{center}
\end{table}

\begin{figure} [H]
\begin{center}
\includegraphics[scale=0.40]{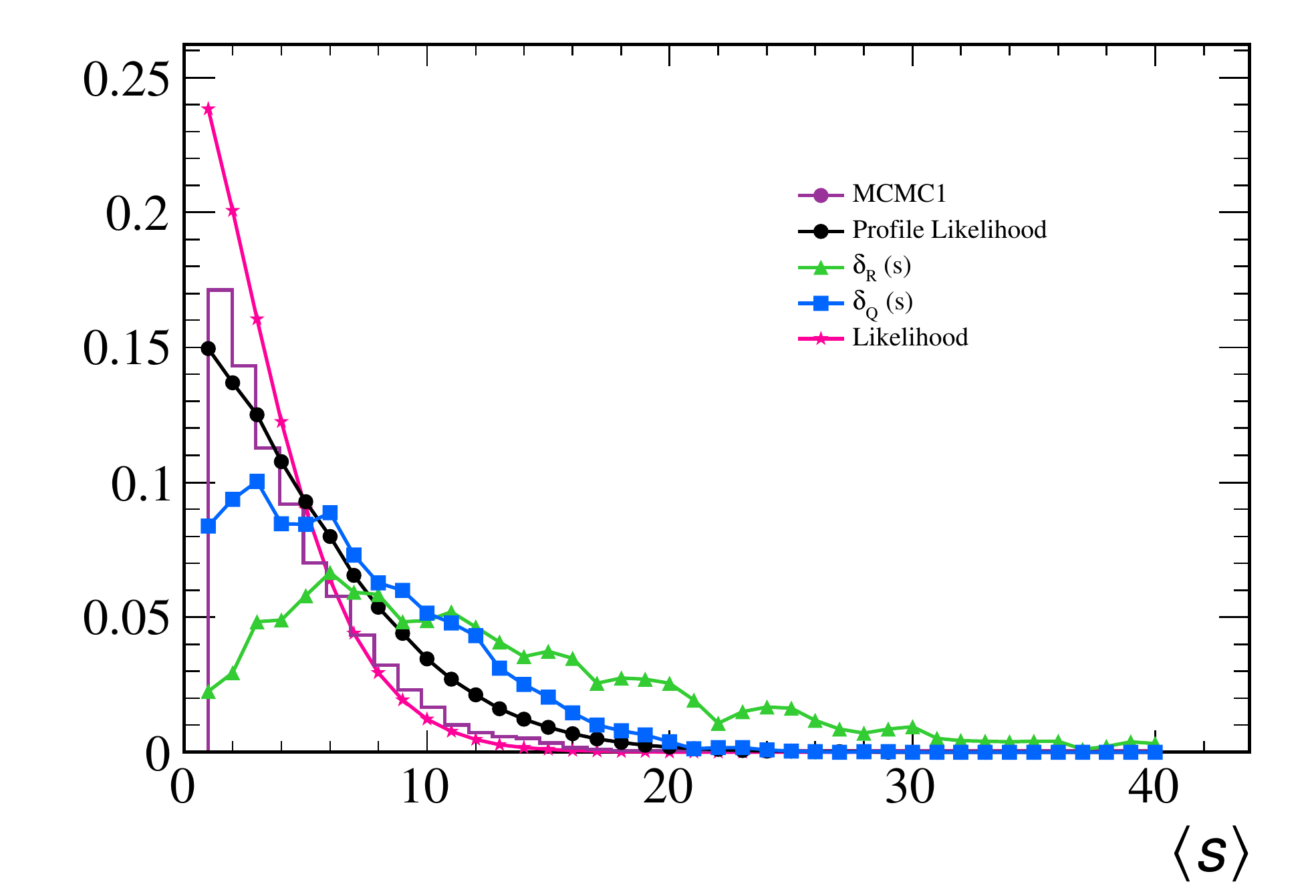}
\includegraphics[scale=0.40]{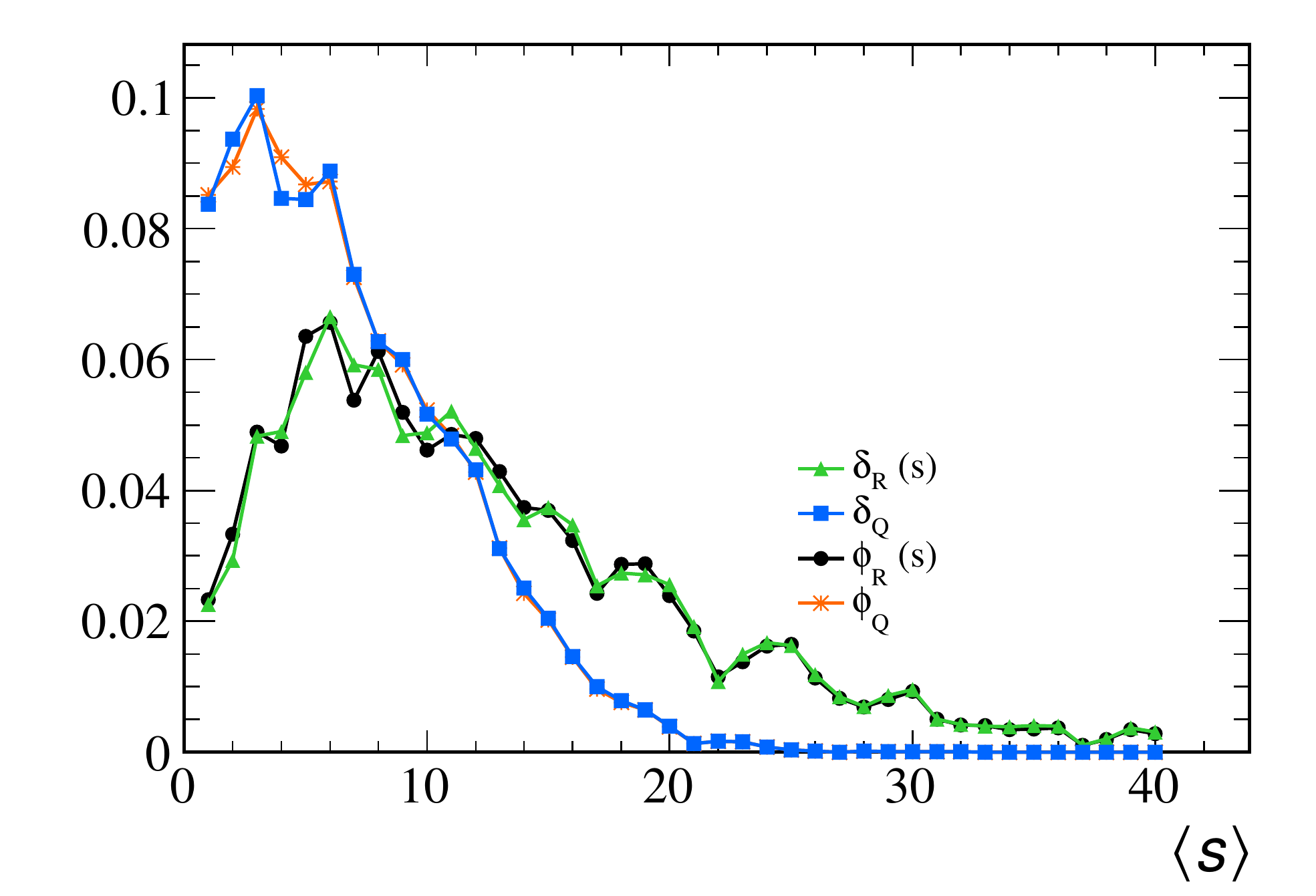}
\includegraphics[scale=0.40]{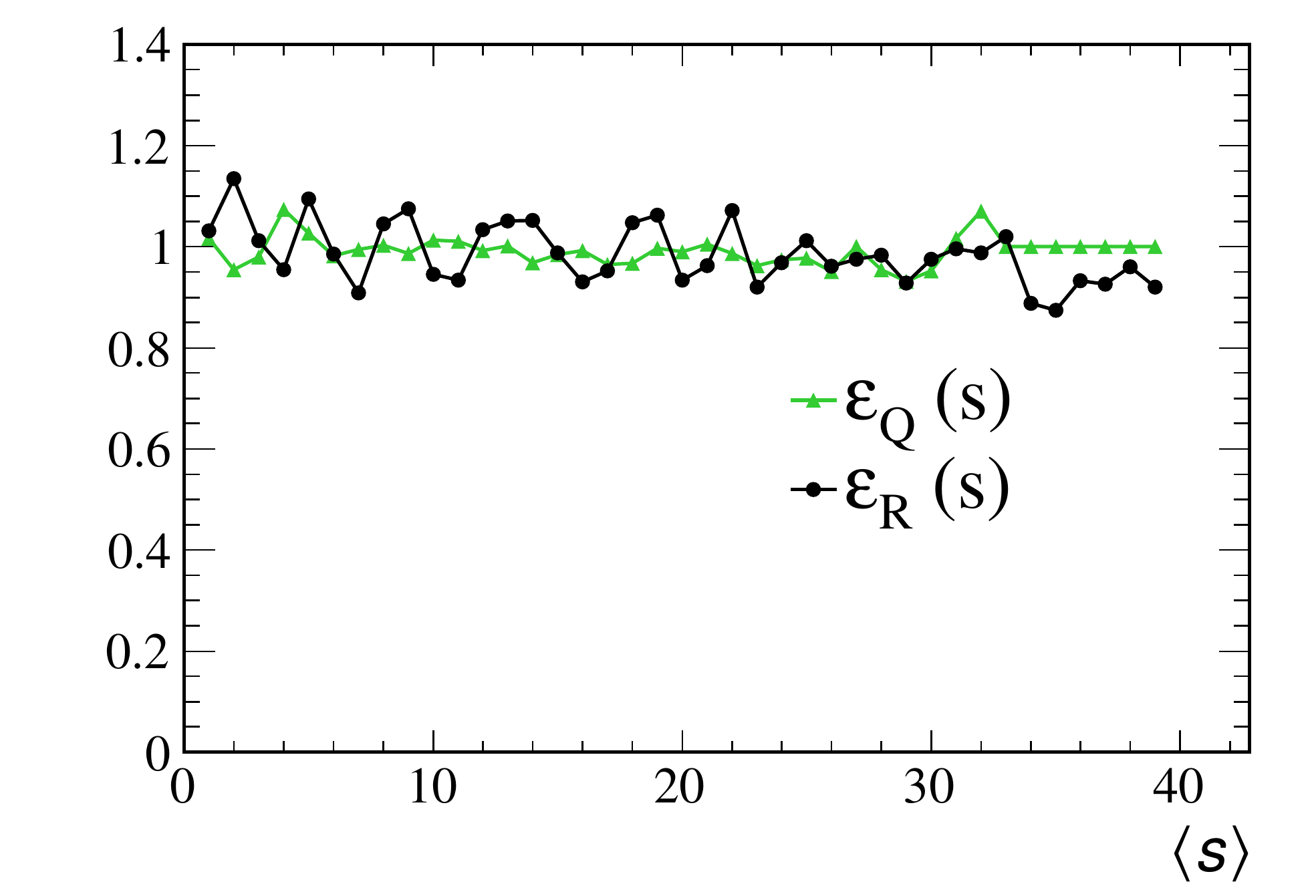}
\caption{{Top: MCM1 (violet solid line), profile likelihood (black circles), $\delta_R(s)$ (green triangles), $\delta_Q(s)$ (cyan squares), likelihood (pink stars). Medium: $\delta_R(s)$ (green triangles), $\delta_Q(s)$ (cyan squares), $\phi_R(s)$ (black circles), $\phi_Q(s)$ (orange asterisks). Bottom: $\varepsilon_Q(s)$ (green triangles), $\varepsilon_R(s)$ (black circles). Experiment with systematics (\secref{ssec:sys}), where $N_{obs} < \left \langle b \right \rangle$.}
\label{fig:dcllack}}
\end{center}
\end{figure}

\begin{figure} [H]
\begin{center}
\includegraphics[scale=0.50]{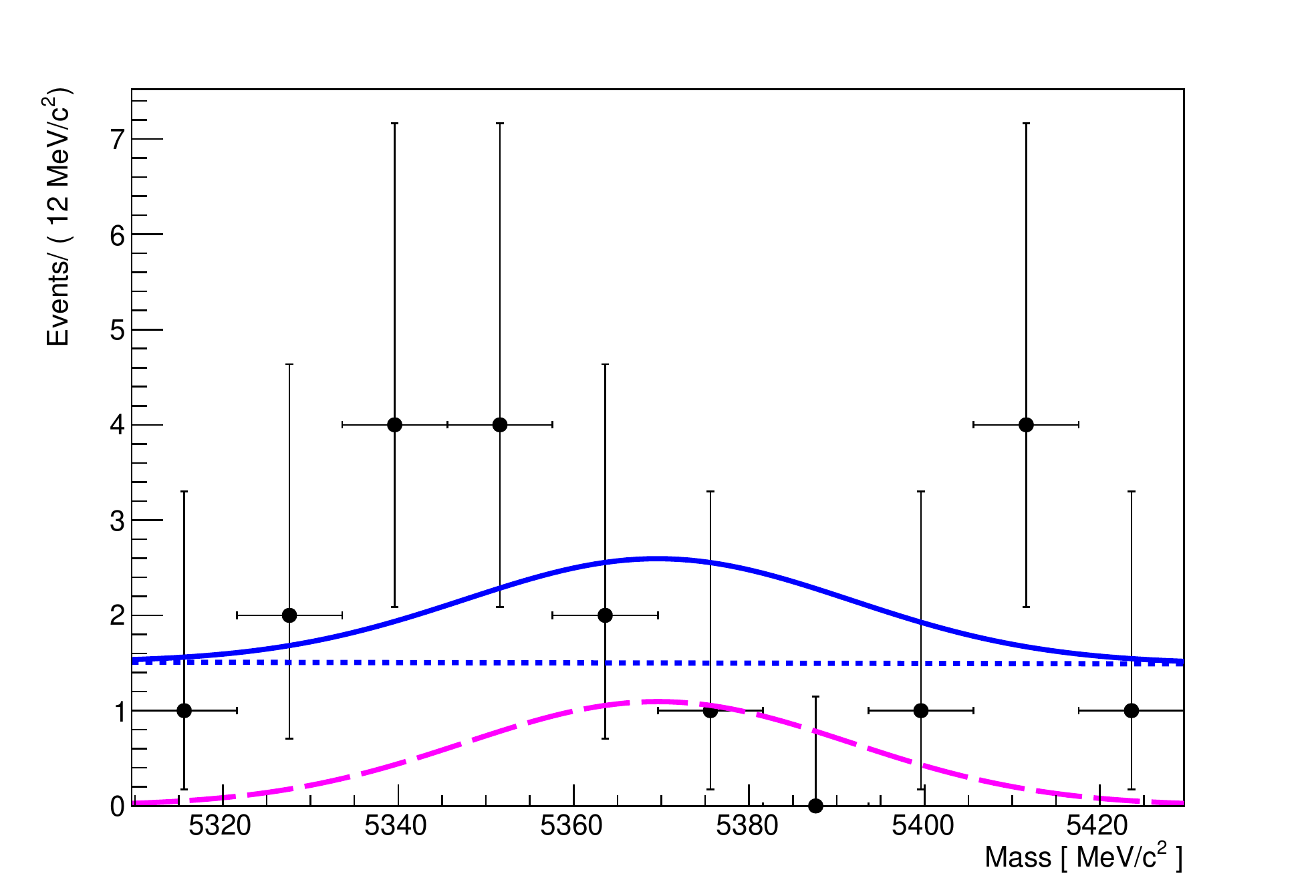}
\caption{{Mass distribution of the generated data for the experiment with systematics (\secref{ssec:sys}), where $N_{obs} < \left \langle b \right \rangle$. The magenta dashed line shows the signal contribution, the dotted blue line the background, and the solid blue line the total model.}
\label{fig:DLL_exp2}}
\end{center}
\end{figure}

{An example where the number of observed events is less than the one predicted with the null hypothesis is also performed, leading to the $\delta(s)$, $\phi(s)$ and $\epsilon(s)$ of \figref{fig:dcllack}}. In \figref{fig:DLL_exp2} we show the mass distribution of the generated data.

\clearpage
\newpage

\section{Conclusions}
\label{sec:conclusions}
In this paper we define the signal derivatives of \CLsb and \CLs and calculate their properties
as estimators of the signal strength. They show similar distributions as obtained from profile likelihood
fits or Markov Chain Monte Carlo routines, and their credible intervals have frequentist coverages.
The functions can be used to construct $\chi^2$ functions for phenomenological analysis as well as for
combinations of experimental results.

\cleardoublepage
\newpage
\section*{Acknowledgements}
 
\noindent We would like to thank T.Junk for helpful discussions in this work, and V. Chobanova for corrections to the draft. We would like to thank financial support from European
Research Council via Grant BSMFLEET 639068 as well as from Xunta de Galicia.

\cleardoublepage
\newpage
\bibliographystyle{JHEP}
\bibliography{main}

\end{document}